\documentclass[%
 reprint,
nofootinbib,
 amsmath,amssymb,
 aps,
]{revtex4-2}

\usepackage{graphicx}
\graphicspath{{./figs/}}
\usepackage{dcolumn}
\usepackage{bm}
\usepackage{hyperref}
\usepackage{braket}
\usepackage{ulem}
\usepackage{comment}

\usepackage{color}
\definecolor{mulberry}{rgb}{0.5,0,0.5}
\definecolor{orange}{rgb}{1,0.65,0}

\newcommand{\cc}[1]{{\color{blue}#1}}

\newcommand{\XiSM}{\ensuremath{\Xi_{\mathrm{SM}}}}

\begin{document}

\hfill MITP-23-026
\vspace{15pt}

\title{Mapping and Probing Froggatt-Nielsen Solutions to the Quark Flavor Puzzle}

\author{Claudia Cornella$^a$}
\email{claudia.cornella@uni-mainz.de}
\author{David Curtin$^b$}
\email{dcurtin@physics.utoronto.ca}
\author{Ethan T.~Neil$^c$}
\email{ethan.neil@colorado.edu}
\author{Jedidiah O. Thompson$^d$}
\email{jedidiah@stanford.edu}

\affiliation{${}^a$ PRISMA$^+$ Cluster of Excellence {\em \&} MITP, 
Johannes Gutenberg University, 55099 Mainz, Germany \\
${}^b$ Department of Physics, University of Toronto, Toronto, ON M5S 1A7, Canada\\
${}^c$Department of Physics, University of Colorado, Boulder, CO 80309, USA\\
${}^d$Stanford Institute for Theoretical Phyiscs, Stanford University, Stanford, CA 94305, USA}

\begin{abstract}
The Froggatt-Nielsen (FN) mechanism is an elegant solution to the flavor problem. In its minimal application to the quark sector, the different quark types and generations have different charges under a $U(1)_X$ flavor symmetry. The SM Yukawa couplings are generated below the flavor breaking scale with hierarchies dictated by the quark charge assignments.
Only a handful of charge assignments are generally considered in the literature. We analyze the complete space of possible charge assignments with $|X_{q_i}| \leq 4$ and perform both a set of Bayesian-inspired numerical scans and an analytical spurion analysis to identify those charge assignments that reliably generate SM-like quark mass and mixing hierarchies. The resulting set of top-20 flavor charge assignments significantly enlarges the viable space of FN models but is still compact enough to enable focused phenomenological study. 
We then apply our numerical methodology to demonstrate that these distinct charge assignments result in the generation of correlated flavor-violating four-quark operators characterized by significantly varied strengths, potentially differing substantially from the possibilities previously explored in the literature.
Future precision measurement of $\Delta F =2 $ observables, along with increasingly accurate SM predictions, may therefore enable us to distinguish among otherwise equally plausible FN charges, thus shedding light on the UV structure of the flavor sector. 
\end{abstract}

\maketitle

\section{Introduction}

The possible origin of the wide and varying hierarchies amongst the quark and lepton masses and mixing angles has long invited speculation.
While such disparate Lagrangian parameters are technically natural for fermions, the fact that the three matter generations appear identical in all other respects 
calls for a dynamical explanation of this so-called flavor puzzle. 

A common strategy is to describe these patterns in terms of an approximate flavor symmetry, a subgroup of $U(3)^5$, whose breaking yields the observed  masses and mixing angles. Historically, the first attempt in this direction was made by Froggatt and Nielsen in 1978 \cite{Froggatt:1978nt}. This strategy, expanded in \cite{Leurer:1993gy,Leurer:1992wg}, became known as the  Froggatt-Nielsen (FN) mechanism (see e.g.\ Ref.~\cite{Altmannshofer:2022aml} for a modern review).
In its simplest form, this mechanism relies on the introduction of a $U(1)_X$ symmetry under which fermions of different generations have different charges. 
This symmetry is broken at some high ``flavor scale'' $\Lambda_F$ by the vacuum expectation value (vev) of a SM-singlet scalar $\phi$, often referred to as the flavon. 
The SM Yukawa couplings are then generated as effective operators suppressed by factors $(\langle \phi \rangle/\Lambda_F)^n = \epsilon^n$, where $\epsilon \lesssim 0.1$ and $n$ is determined by the $U(1)_X$ charges of the Higgs and the respective fermions. 
Assuming that these operators are generated in the UV completion of the model through 
heavy particle exchange with presumably $\mathcal{O}(1)$ coefficients
(see e.g.\ Refs.~\cite{Calibbi:2012yj, Alonso:2018bcg, Bonnefoy:2019lsn}), 
the hierarchies of the Yukawa couplings in the flavor basis  are generated by the $U(1)_X$ charge assignments of the SM fields and the flavon vev relative to the UV scale, $\epsilon$.

FN models have been extremely well studied over the last decades (see e.g.\ Refs. \cite{Leurer:1992wg,Leurer:1993gy,Ibanez:1994ig,Jain:1994hd,Dudas:1995yu, Dudas:1996fe, Irges:1998ax,Calibbi:2012yj,Calibbi:2016hwq,Ema:2016ops,Bauer:2016rxs, Dery:2016fyj, Alonso:2018bcg, Alanne:2018fns, Smolkovic:2019jow, Bonnefoy:2019lsn, Bordone:2019uzc, Fedele:2020fvh,Altmannshofer:2022aml}), but curiously, the vast literature only considered very few choices of the flavor symmetry charge assignments for the SM fields. 
The studied scenarios typically identify $\epsilon$ with the Cabibbo angle of the Cabibbo-Kobayashi-Maskawa (CKM) matrix and perform a spurion analysis to find a solution for the required fermion charges. 
The selection of these charges carries important phenomenological implications, since FN models have a variety of experimental signatures at low energies -- most obviously various flavor-changing processes -- the details of which depend on the exact charge assignment chosen.

In FN models of the quark sector, flavor-changing neutral current (FCNC) constraints involving the light generations bound the flavor scale to be above $\Lambda_F \gtrsim 10-100$~PeV~\cite{Altmannshofer:2022aml}. This makes FN models (like most solutions to the flavor puzzle relying on a single flavor scale $\Lambda_F$) notoriously hard to probe experimentally. 
However, the next decades may see significant advances in our ability to probe flavor violation beyond the SM, whether with new data from on-going experiments (most notably LHCb and Belle II), concrete proposals for future high-energy colliders \cite{Agapov:2022bhm, Gao:2022lew, ILCInternationalDevelopmentTeam:2022izu, Bernardi:2022hny, CEPCStudyGroup:2018ghi, Brunner:2022usy, MuonCollider:2022xlm, Aime:2022flm}, the hypothetical possibility of future flavor factories, or theoretical advances to improve the precision of SM predictions for experimentally well-measured processes. 
It is thus pertinent to investigate whether experimental insights can be gained regarding FN models in the foreseeable future, despite their characteristic high energy scale.  

In this paper, we perform the first step in such a model-exhaustive program by identifying the most general ``theory space'' of natural FN models with a single Higgs field that can address the SM flavor problem in the quark sector.
As pointed out in Ref.~\cite{Fedele:2020fvh}, flavor charge assignments beyond the few canonical choices can generate close matches to the SM, seemingly with natural $\mathcal{O}(1)$ choices for the various coefficients. 
Inspired by this observation, our analysis considers all possible flavor charge assignments up to $|X_q| \leq 4$ in a FN setup restricted to the quark sector, then performs numerical scans over natural $\mathcal{O}(1)$ choices of all the Yukawa coefficients, to determine in a maximally agnostic and general fashion which charge assignments and flavon vevs can generate ``SM-like'' quark masses and mixings. 
We find that the space of fully natural FN models is much larger than previously understood in the literature, including at least several dozen models depending on the precise interpretation of our results.

Unlike previous such analyses, we select the ``SM-like'' choices in a Bayesian-inspired fashion without explicitly fitting to the SM. We believe this more accurately reflects the intended spirit of the FN solution, that the SM hierarchies should emerge ``naturally.''\footnote{Our approach to quantifying flavour naturalness is very similar to~\cite{Aloni:2021wzk}, though that study focused on lepton CP violation in the MSSM for a handful of natural charge assignments. For an example of applying more strictly Bayesian analyses to select FN models of the lepton sector, see~\cite{Altarelli:2002sg, Altarelli:2012ia, Bergstrom:2014owa}.} Reassuringly, we do also find that the SM-like FN setups easily yield many stable exact fits to the SM, and that our numerical results can be understood in the context of an analytical spurion analysis that assumes small rotations between the flavor- and mass-bases.

With this natural theory space of viable FN models now defined, we study its phenomenology by estimating the size of various flavor-violating observables, most notably meson-antimeson mixing,  in each model and comparing the range of predictions to current constraints. This analysis shows that different charge assignments or “textures” yield distinct predictions for flavor-violating observables, and highlights the specific observables that could be most effective in distinguishing between different FN solutions to the flavor puzzle in the future. 

We hope that this kind of analysis can be helpful for future studies of flavor physics, including SMEFT global fits, by providing a ``theoretical lamp post'' that can direct attention on the most well-motivated parts of the vast landscape of FN models and flavor observables.

This paper is structured as follows. In Section~\ref{s.review} we briefly review the Froggatt-Nielsen mechanism. The numerical analysis of all possible FN models with $|X_q| \leq 4$ is presented in Section~\ref{s.numerical}. The results of this analysis are corroborated by an analytical spurion analysis in Section~\ref{s.analytical}. he phenomenological implications are explored in Section~\ref{s.pheno}, and we conclude in Section~\ref{s.conslusions}. Details on SM fits, a custom tuning measure appropriate for FN models, and a variety of necessary statistical checks are collected in the Appendices.

\section{Review of the Froggatt-Nielsen mechanism}

\label{s.review}
In this Section we briefly review the Froggatt-Nielsen mechanism in its minimal form. 
Our treatment and notation parallels closely that of Ref.~\cite{Fedele:2020fvh}.
As stated in the introduction, the FN mechanism consists of adding a new $U(1)_X$ global flavor symmetry to the SM, along with a scalar field $\phi$ whose vev spontaneously breaks this symmetry. Without loss of generality we take $X_\phi = 1, X_H = 0$ and the vev of $\phi$ to be in the positive real direction, $\braket{\phi} = \braket{\phi}^\dagger = \epsilon \Lambda_F$, where $\Lambda_F$ is the characteristic scale of spontaneous symmetry breaking. SM matter fields are then assigned $U(1)_X$ charges, forcing the scalar to appear in the SM Yukawa couplings in order to preserve the symmetry. We denote the left-handed quark doublets as $Q_i$ and the right-handed up- and down-quark singlets $u_i$ and $d_i$, where $i = 1, 2, 3$ is an index running over the fermion families. The quark mass terms in the low-energy SM effective field theory then take the form:
\begin{align} \label{eq:FNYukawa}
\mathcal{L}_{\mathrm{Y}} &\supset - c_{i j}^u \bar{Q}_i \widetilde{H} u_j \epsilon^{|X_{Q_i} - X_{u_j}|} 
     - c_{i j}^d \bar{Q}_i H d_j \epsilon^{|X_{Q_i} - X_{d_j}|} + \mathrm{h.c.}
\end{align}
where $H$ is the SM Higgs doublet, and $c^{u,d}$ are arbitrary $3 \times 3$ complex matrices with (presumably) $\mathcal{O}(1)$ entries. 
Comparing Eq.~\ref{eq:FNYukawa} to the usual SM Yukawa Lagrangian
\begin{equation}
    \mathcal{L}_{\mathrm{SM}} \supset - Y_{i j}^u \bar{Q}_i \widetilde{H} u_j - Y_{i j}^d \bar{Q}_i H d_j + \mathrm{h.c.}\,,
\end{equation}
we can read off the Yukawa matrices in terms of the FN parameters: 
\begin{align} \label{eq:YukawaMatrixInTermsOfCEpsilon}
    Y^{u}_{i j} = c^{u}_{i j} \epsilon^{n^u_{ij}}\,, && 
     Y^{d}_{i j}  = c^{d}_{i j} \epsilon^{n^d_{ij}}\, , 
\end{align}
where
\begin{align} \label{eq:nund}
    n^u_{i j} = |X_{Q_i} - X_{u_j}|\,, && 
    n^d_{ij} = |X_{Q_i} - X_{d_j}| \, . 
\end{align}
From this expression it is clear that, if $\epsilon \ll 1$, the SM Yukawa couplings can exhibit large hierarchies even if all entries of the coefficient matrices $c^{u,d}$ are $\mathcal{O}(1)$.
These can be related to observable quantities -- the quark masses and mixing angles -- by performing unitary flavor rotations of the quark fields.

\section{Systematic exploration of Froggatt-Nielsen Models}

\label{s.numerical}

We consider all possible charge assignments with $|X| \leq 4$ for all SM quarks. Since $y_t \approx 1$, baryon number is conserved, and permutations of the quark fields do not constitute a physical difference between FN models, we can set $X_{q_3}=X_{u_3}=0$ and adopt the ordering convention that $|X_{Q_i}| \geq |X_{Q_j}|$ for $i < j$ when all $X_Q$ have the same sign, otherwise $X_{Q_i} \geq X_{Q_j}$. (Similarly for $X_u$, or $X_d$.)
In addition, following \cite{Fedele:2020fvh} we remove ``mirror'' charges which are related by multiplying all of the charges $X$ by $-1$ (and then enforcing the above ordering convention).  This gives a total of 167,125  charge assignments that are physically inequivalent in the IR.

We now wish to assess these textures based on how ``typical'' it is for them to reproduce the flavor hierarchies of the SM.  
Given a certain charge assignment $\boldsymbol{X}=\left \{ X_Q, X_u, X_d \right\}$, we randomly draw some large number of $\mathcal{O}(1)$ complex coefficient matrices $c^{u,d}$ (details discussed below) and calculate the SM parameters for each instance of $c^{u,d}$ as a function of $\epsilon$. 
For a given choice of $\epsilon$, 
we can then calculate the observable with the maximum fractional deviation from its SM value:
\begin{equation} \label{eq:defXi}
\XiSM = \max_{i} \exp \Bigg| \ln \left( \frac{\mu_i^{\text{guess}}}{\mu_i^{\text{SM}}} \right) \Bigg| \ ,
\end{equation}
where $\mu_i$ ranges over the six quark masses, the three independent CKM entries $|V_{12}|, |V_{13}|$, and $|V_{23}|$, and the absolute value of the Jarlskog invariant $|J|$.  The precise numerical values we use can be found in Table~\ref{tab:SMparams} in Appendix~\ref{app:prior}.
Since $\epsilon$ plays the central role of generating the overall hierarchy,
we minimize $\XiSM$ with respect to $\epsilon$ for each instance of the $c^{u,d}$ coefficient matrices to match as closely as possible the SM values.

\begin{table}[]
    \centering
    \resizebox{\columnwidth}{!}{\begin{tabular}{c|c|c|c|c|c|c|c|c|c|c}

        Num. & $X_{Q_1}$ & $X_{Q_2}$ & $X_{u_1}$ & $X_{u_2}$ & $X_{d_1}$ & $X_{d_2}$ & $X_{d_3}$ & $\mathcal{F}_2 \; (\%)$ & $\mathcal{F}_5 \; (\%)$ & $\epsilon$ \\
         \hline
         1 & 3 & 2 & -4 & -2 & -3 & -3 & -3 & 2.7 & 67 & 0.17 \\
2 & 3 & 2 & -4 & -2 & -4 & -3 & -3 & 2.5 & 66 & 0.18 \\
3 & 3 & 2 & -3 & -1 & -3 & -2 & -2 & 1.9 & 56 & 0.12 \\
4 & 3 & 2 & -4 & -1 & -3 & -3 & -3 & 1.5 & 65 & 0.16 \\
5 & 4 & 3 & -4 & -2 & -4 & -3 & -3 & 1.2 & 52 & 0.23 \\
6 & 3 & 2 & -4 & -1 & -3 & -3 & -2 & 1.1 & 63 & 0.15 \\
7 & 4 & 2 & -4 & -2 & -4 & -3 & -3 & 1.1 & 47 & 0.21 \\
8 & 3 & 2 & -3 & -1 & -2 & -2 & -2 & 0.9 & 41 & 0.11 \\
9 & 3 & 2 & -3 & -1 & -3 & -3 & -2 & 0.9 & 55 & 0.14 \\
10 & 3 & 2 & -4 & -2 & -3 & -3 & -2 & 0.9 & 59 & 0.16 \\
11 & 2 & 1 & -3 & -1 & -2 & -2 & -2 & 0.8 & 52 & 0.06 \\
12 & 4 & 3 & -4 & -1 & -4 & -3 & -3 & 0.8 & 52 & 0.22 \\
13 & 4 & 3 & -4 & -2 & -4 & -4 & -3 & 0.8 & 50 & 0.24 \\
14 & 3 & 2 & -4 & -2 & -4 & -3 & -2 & 0.7 & 56 & 0.17 \\
15 & 4 & 3 & -4 & -2 & -3 & -3 & -3 & 0.7 & 43 & 0.22 \\
16 & 4 & 3 & -4 & -1 & -3 & -3 & -3 & 0.6 & 45 & 0.21 \\
17 & 4 & 2 & -4 & -2 & -4 & -4 & -3 & 0.6 & 48 & 0.22 \\
18 & 4 & 2 & -4 & -2 & -3 & -3 & -3 & 0.6 & 38 & 0.2 \\
19 & 3 & 2 & -4 & -1 & -3 & -2 & -2 & 0.6 & 56 & 0.14 \\
20 & 3 & 2 & -3 & -2 & -3 & -3 & -2 & 0.6 & 45 & 0.15
    \end{tabular}}
    \caption{
    The top 20 textures ranked by the fraction of the time that a random draw of $c^{u,d}$ coefficient matrices result in all quark masses, CKM matrix elements, and the Jarlskog invariant within a factor of $2$ of the SM (allowing $\epsilon$ to vary to minimize deviation from the SM).  For each texture, $\mathcal{F}_2$ ($\mathcal{F}_5)$ is the fraction within $2$ (5), and $\epsilon$ is the average value of $\epsilon$ for those coefficients that end up within a factor of $2$. For all of these textures, $X_{Q_3} = X_{u_3} = 0$.}
    \label{tab:bestWithin2}
\end{table}

After repeating this process for  each charge assignment $\boldsymbol{X}$, 
we define $\mathcal{F}_2$ ($\mathcal{F}_5$) as the fraction of random coefficient choices that yields $\XiSM \leq 2$~$(5)$ for each model. This allows us to define ``global naturalness criteria'' (i.e. independent of a particular \textit{fit} to the SM) for a given texture to be a good candidate for reproducing the SM hierarchies, namely requiring that $\mathcal{F}_2$ or $\mathcal{F}_5$ be above some lower bound. 

Out of the total 167,125 possible charge assignments, we only find about 10 for which $\mathcal{F}_2 \gtrsim 1\%$.
We adopt the latter criterion as the most stringent measure of naturalness for a texture to solve the SM quark flavor problem, and show the top-20 possible charge assignemnts ranked by $\mathcal{F}_2$ in Table~\ref{tab:bestWithin2}.We also show the average value of $\epsilon$ for each texture. 
The distribution of preferred $\epsilon$ values is very narrow, indicating that each texture has a uniquely suited value of $\langle \phi \rangle/\Lambda_F$ to reproduce the SM.
Some of these good textures correspond to those identified already in the literature, but, to the best of our knowledge, most of them have not been considered before.\footnote{Texture 3 first appeared in the seminal papers~\cite{Leurer:1992wg,Leurer:1993gy}.}

This is quite remarkable, given how similarly they all naturally generate SM-like mass and mixing hierarchies. 
Textures where all quark masses and CKM entries lie within a factor of 5 of their measured SM values are even more common, with about 430 textures satisfying $\mathcal{F}_5 \gtrsim 10\%$. This constitutes a large collection of textures that are well-motivated in this model-independent way. The complete ranking of these textures is included in the file \texttt{charges.csv} provided in the auxiliary material.\footnote{Note that because the auxiliary file is rank ordered by $\mathcal{F}_5$, it does not match the order given in Table~\ref{tab:bestWithin2}, and the precise ordering of textures with very similar values of $\mathcal{F}_5$ (or $\mathcal{F}_2$) is subject to small numerical fluctuations. Several textures in this larger list have been studied in the past, for example~\cite{Calibbi:2015sfa} examines a texture with $\mathcal{F}_{5(2)} = 33 (0.1)\%$.}

\begin{figure}[]
    \centering
    \includegraphics[width=\linewidth]{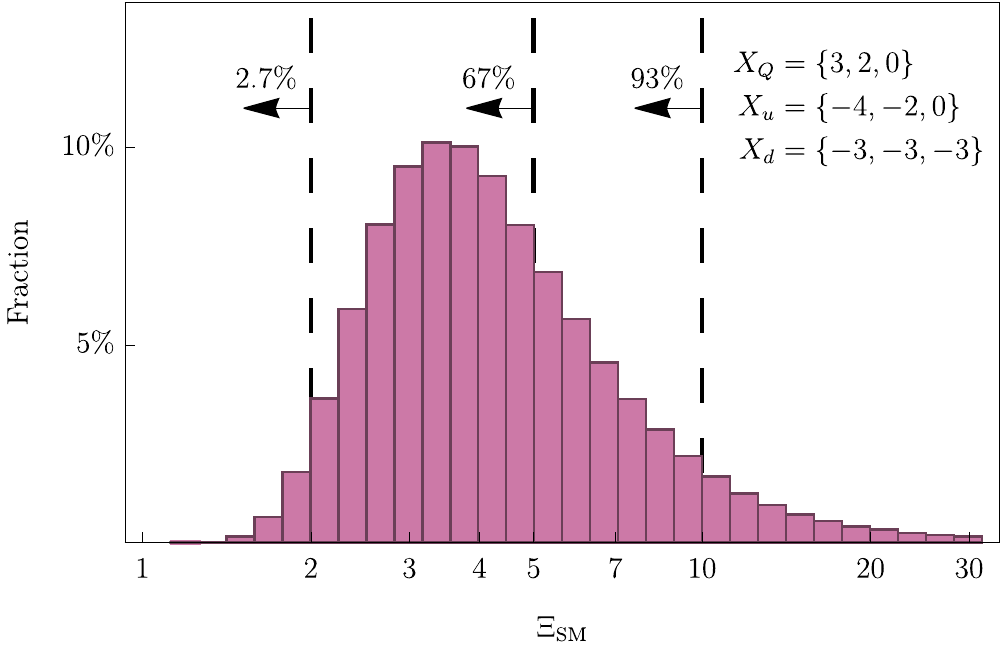}
    \caption{A probability distribution function for $\XiSM$ for the top-ranked texture in Tab.~\ref{tab:bestWithin2}. To produce this plot we randomly generate many different choices of $c^{u/d}$ matrices and compute $\XiSM$ for each one (allowing $\epsilon$ to vary to minimize deviation from the SM). Also shown are the total fraction of coefficient matrices that result in $\XiSM < 2, 5, \text{and } 10$ respectively. For this texture, typical random $\mathcal{O}(1)$ matrices of coefficients will result in all quark masses and CKM parameters within a factor of $\approx 4$ of their measured SM values.}
    \label{fig:best2}
\end{figure}
As mentioned above, to understand how close a texture generically is to the SM, we may look at the distribution of $\XiSM$ values for a given texture over many different random choices of the $c^{u,d}$ matrices.  In Fig.~\ref{fig:best2}, we show such a distribution for  the top texture from Table~\ref{tab:bestWithin2}.  As we can see, the distribution is clustered around $\XiSM \sim 4$ with broad tails on either side, indicating that this texture generically results in physical parameters with the proper SM-like hierarchies regardless of the $\mathcal{O}(1)$ coefficients $c^{u,d}$.  Of course getting the precise SM values requires a precise choice of these coefficients, but the hierarchies themselves are robust.

\begin{figure}[]
    \centering
    \includegraphics[width=\linewidth]{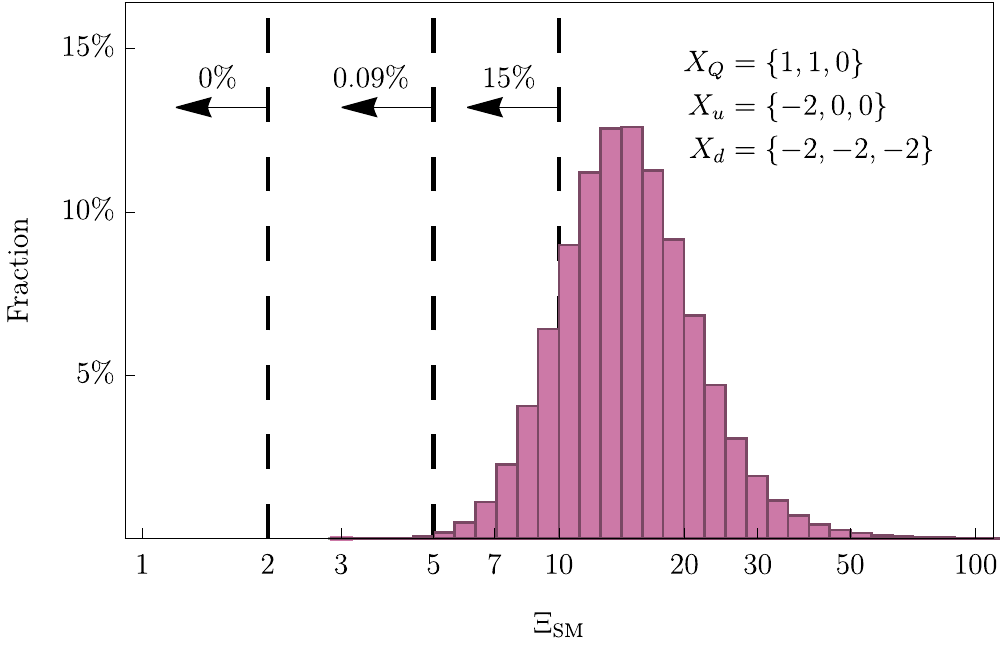}
    \caption{Similar to Fig.~\ref{fig:best2} for a texture that does not perform well by our metric. This texture does have a technically natural fit to the SM (as found in Ref.~\cite{Fedele:2020fvh}), but typical $\mathcal{O}(1)$ coefficient matrices for this texture result in deviations of more than an order of magnitude from the measured SM parameters.}
    \label{fig:badcharge}
\end{figure}

We may also look at such a distribution for a charge assignment that does not perform well by this metric.  This category includes many textures previously mentioned in the literature, including for example most of the textures listed in Tables 1 and 2 of Ref.~\cite{Fedele:2020fvh}.  The third texture in Table 2 of \cite{Fedele:2020fvh} results in the distribution for $\XiSM$ shown in Fig.~\ref{fig:badcharge}.  It is quite reasonable to ask why these textures show up as good in that analysis but not ours, and the answer is simple: Ref.~\cite{Fedele:2020fvh} searches for textures for which there is at least one technically natural choice of $c^{u,d}$ coefficients that approximately reproduces the SM, whereas we look for textures which generically produce SM-like hierarchies.  With random $\mathcal{O}(1)$ coefficient matrices, a texture like that of Fig.~\ref{fig:badcharge} results in a typical deviation from SM parameters of more than an order of magnitude, but there does exist a very special choice of $c^{u,d}\sim \mathcal{O}(1)$ that results in the SM.  This choice of coefficients may be technically natural 
but it is still exceedingly rare, as Fig.~\ref{fig:badcharge} shows.  The difference thus stems from our ``global'' choice of metric for naturalness, which prioritizes textures that can lead to SM-like parameters without need for additional dynamics or precise restrictions on the coefficients.

To verify that our global naturalness criterion also produces viable and ``locally natural'' solutions to the SM flavor problem, we also have to check that each of the textures in Table~\ref{tab:bestWithin2}
yields exact and untuned numerical fits to the SM, including uncertainty on the SM parameters.  This necessitates the definition of a custom tuning measure that is more appropriate for FN models than standard choices like the Barbieri-Giudice measure~\cite{Ellis:1986yg, Barbieri:1987fn}, taking into account that any change in the UV theory would likely perturb all coefficients in Eq.~\ref{eq:FNYukawa} at once, rather than one at a time. The details are in Appendix~\ref{app:smfits}, but the upshot is that each of the textures in Table~\ref{tab:bestWithin2} readily yields many good SM fits that are completely untuned with respect to simultaneous random uncorrelated perturbations of all coupling coefficients.  This confirms that our global naturalness criterion guarantees particular technically natural solutions to the SM flavor problem as well.

We perform several statistical checks to make sure our conclusions are robust. 
The results obtained here depend on the details of the statistical distributions from which we draw the coefficients in the $c^{u,d}$ matrices.  Taking a Bayesian-inspired point of view, these distributions can be thought of as prior probability distributions  for the $c^{u,d}$ coefficients.  For the analysis shown here, we use a ``log-normal'' prior in which the logarithm of the magnitude of each $c^{u,d}$ entry is drawn from a Gaussian distribution.  We have repeated the same analysis for a wider log-normal distribution, and a third ``uniform'' prior, in which the real and imaginary parts of each $c^{u,d}$ are drawn uniformly from the range $[-3,3]$. Up to modest reorderings in our ranked list of top textures, our results are robust with respect to changing the prior. We also confirm that the SM-like hierarchies for our good textures are robustly determined by the flavor charges and the small $\epsilon$ rather than anomalous hierarchies amongst the randomly drawn $c^{u,d}$ coefficients, and that individual SM observables are roughly uncorrelated and drawn from a distribution roughly centered on the correct SM value for each good texture. For details, see Appendix~\ref{app:prior}.

\section{Analytical Spurion Analysis}
\label{s.analytical}

Our numerical study identified many new plausible FN textures beyond those that have been studied in the literature. It would be useful to understand analytically why the textures in Table~\ref{tab:bestWithin2} reliably yield SM-like hierarchies.
If viable FN models involve no large rotations between the flavor- and the mass-basis, then it is possible to estimate masses and mixings via an analytical spurion analysis. We can easily test this assumption.

In general, the flavor-basis Yukawa matrix $Y^u$ (identically for $Y^d$) can be related to the diagonal mass-basis Yukawa $\hat Y^u$ via the usual singular value decomposition $Y^u = U_u \hat Y^u W_u^\dagger$. Under the near-aligned assumption, and adopting the strict ordering of charges defined at the beginning of Section~\ref{s.numerical}, the  mass-basis Yukawa couplings are simply given by
\begin{equation}
    \label{e.Yspurion}
    \hat Y^u_{ij} \sim \epsilon^{n^u_{ii}} \delta_{ij}
\end{equation}
where the exponent matrices $n^{u,d}$ are defined terms of the $X_{Q,u,d}$ charges in Eq.~\ref{eq:nund}.
Under the same strict assumptions,\footnote{Eq.~\ref{e.UW} does not apply for flavor charge assignments where the heaviest up and down quarks are not in the same generation, but those are irrelevant for our analysis.} the magnitudes of the elements of the rotation matrices can be written as
\begin{eqnarray}
\nonumber 
    (U_u)_{ij} &\sim& \left\{
    \begin{array}{lll}
        1 &  
        &  i = j\\
        \epsilon^{n^u_{ij} - n^u_{jj}} & 
        & i < j \\
        \epsilon^{n^u_{ji} - n^u_{ii}} & 
        & i > j
    \end{array}
    \right.
\\
\label{e.UW}
\\ 
\nonumber 
(W_u)_{ij} &\sim& \left\{
    \begin{array}{lll}
        1 &  
        &  i = j\\
        \epsilon^{n^u_{ji} - n^u_{jj}} & 
        & i < j \\
        \epsilon^{n^u_{ij} - n^u_{ii}} & 
        & i > j
    \end{array}
    \right. \ .
\end{eqnarray}
This makes it straightforward to obtain magnitude estimates for the elements of $V_\mathrm{CKM} = U_u^\dagger U_d$.\footnote{
For about half of the textures in Table~\ref{fig:best2}, this yields $V_\mathrm{CKM}$ estimates that follow the Wolfenstein parameterization (with $\lambda \rightarrow \epsilon$), while the other half show slight deviations from this pattern.}

For a given FN charge assignment and choice of $\epsilon$, we can now compute the spurion estimate  $\mu_i^{\text{spurion}}$ for $i = $ each of the six quark masses and $|V_\mathrm{CKM}|_{12,23,13}$, using Eqns.~\ref{e.Yspurion},~\ref{e.UW} verbatim, working to lowest order in $\epsilon$, and setting all coefficients to 1. 
To assess how SM-like a given FN charge assignment is, we consider the logs of the ratios by which these estimates deviate from the SM, added in quadrature, 
\begin{equation} \label{eq:d}
d = \left[ \sum_i \log_{10}^2 \left( \frac{\mu_i^{\text{spurion}}}{\mu_i^{\text{SM}}} \right) \right]^{1/2} \ ,
\end{equation}
with $\epsilon$ chosen to minimize $d$.
This measure always takes all SM observables into account and is more appropriate for the approximate nature of the spurion analysis than Eq.~\ref{eq:defXi}.
We would expect that FN textures that naturally give a good fit to the SM would satisfy $d \lesssim \mathcal{O}(1)$.

Indeed, we find that all the textures in Table~\ref{fig:best2} satisfy $d < 2.4$.
Of all 167,125 possible charge assignments, there are only 36 additional possibilities that also satisfy this criterion, and all of them are in the top 160 of our ranked list of textures obtained in the numerical analysis of the previous section. This analytical spurion estimate is therefore fully consistent with our full numerical study, which
gives us additional confidence that the global naturalness criterion derived in our numerical study is theoretically plausible. 
Furthermore, this convergent result confirms the expectation that natural FN models always feature a high degree of alignment between the flavor and the mass bases.\footnote{This was confirmed in the numerical analysis, where coefficient choices with small values of $\XiSM$ for textures in Table~\ref{tab:bestWithin2} always feature small mixing angles in $U_{u,d}, W_{u,d}$.}

Obviously, the specific $d < 2.4$ criterion was derived \textit{a posteriori} from the results of the numerical study, but if one were to guess at a reasonable upper bound for $d$ that natural SM-like FN models must satisfy, a number like $d < 3$ might plausibly come to mind, which yields a still very modest total of 149 textures (including the top 20).
As with any such definition, the cutoff for what constitutes a natural model is somewhat arbitrary, and if one wants to make quantitative statements the fully numerical approach of the previous section is necessary.
However, it is interesting to note that the numerical study was not necessary to make the much more important qualitative observation that there are many different FN charge assignments, of the order of several dozen at least, that very naturally give SM-like hierarchies. The crucial ingredient is merely to formulate a well-defined and general criterion and check it across all possible flavor charges.

\section{Phenomenological Implications}
\label{s.pheno}

\begin{figure*}
\begin{center}
\includegraphics[width=1\linewidth]{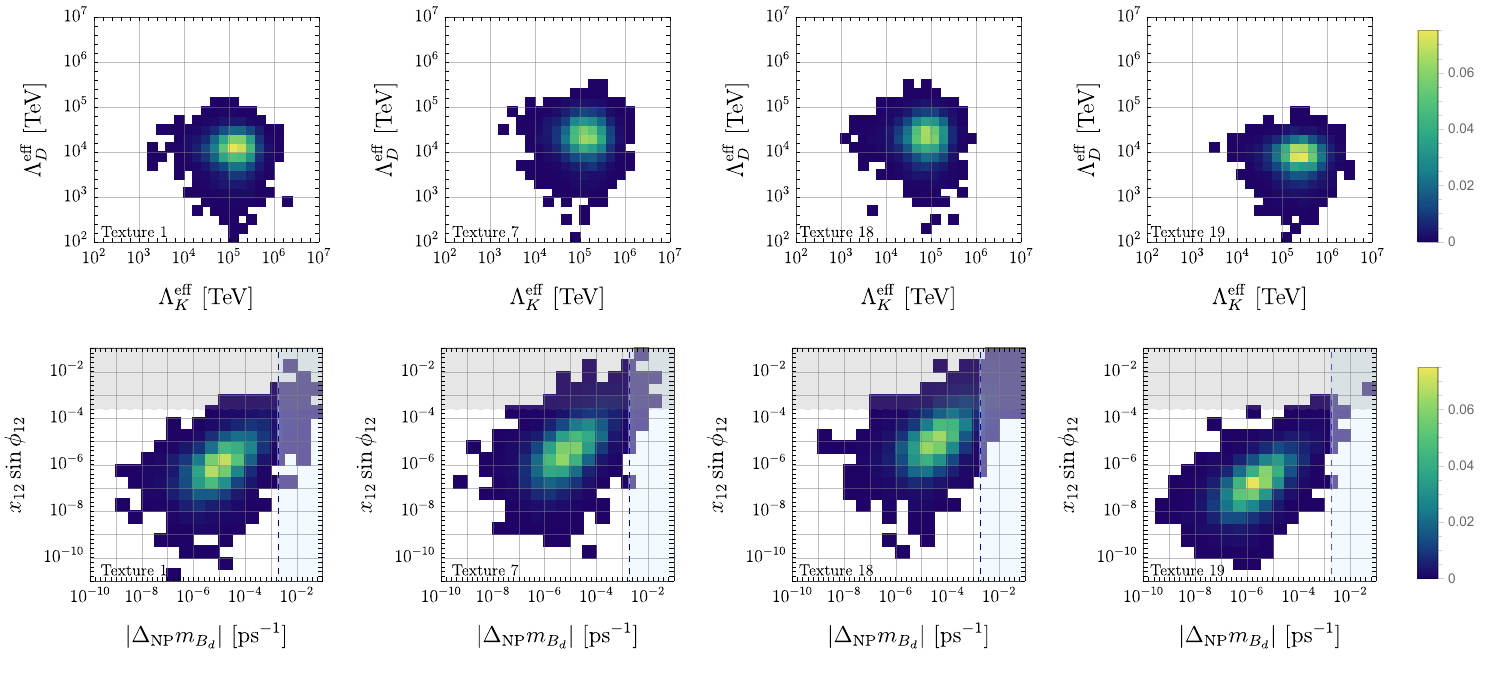}\\

    \caption{
    \textit{Top row.} Distribution of $(\Lambda^{\rm eff}_K, \Lambda^{\rm eff}_D)$ for representative textures 1, 7, 18, and 19 from Table~\ref{tab:bestWithin2}, obtained by generating $10^4$ different natural SM fits per texture (see text). $\Lambda^{\rm eff}_K$ and $\Lambda^{\rm eff}_D$ are the values of $\Lambda_F$ that saturate the bound on $\epsilon_K$ from $K-\bar K$ mixing and $x_{12} \sin \phi_{12}$ from $D - \bar D$ mixing, respectively.  The coloring indicates the fraction of points in a single bin, as explained in the bar legend on the right. \textit{Bottom row}. Correlation between $|\Delta_{\rm NP} m_{B_d}|$ from \cc{$B_d - \bar B_d$} mixing and $x_{12} \sin \phi_{12}$ for the same textures, obtained by setting $\Lambda_F = \Lambda^{\rm eff}_K$ for each point. The shaded gray band represents points that are already excluded by bounds on $D - \bar D$ mixing. The light-blue dashed line shows the current experimental precision on $\Delta m_{B_d}$ \cite{ParticleDataGroup:2024cfk}. 
    }
    \label{fig:phenocorrelations}
\end{center}
\end{figure*}
Flavor-violating effects are the most relevant experimental signatures of FN models. At energies well below $\Lambda_F$ these can be effectively parameterized within the framework of Standard Model Effective Field Theory (SMEFT). In the ``FN flavor basis'' of Eq.~\ref{eq:YukawaMatrixInTermsOfCEpsilon}, where flavor charges are well-defined for each quark generation, the couplings of flavor-changing operators are suppressed by powers of $\epsilon^n$, where the value of $n$ is set by the charges of the fields appearing in the operator. For example, for a generic 4-quark operator in the SMEFT~\footnote{For the sake of simplicity, Dirac structures as well as Lorentz and gauge group indices are left implicit. Upper-case roman indices $A$ are taken to label the different possible dimension-6 flavor-violating quark operators. Flavor indices are indicated explicitly.}   
\begin{equation}
    \mathcal{O}^A_{ijkl} = \frac{1}{(\Lambda^A_{\mathrm{eff}, ijkl})^2} (\bar q_i q_j)(\bar q_k q_l)
\end{equation}
where $i,j,k,l$ denote different quark flavors, and $q$ can represent either $Q, d$ or $u$ type quarks, 
we expect the associated effective scale to be given by
\begin{equation}
\label{e.LambdaEffFNbasis}
\frac{1}{(\Lambda^A_{\mathrm{eff}, ijkl})^2} = c^{A}_{ijkl} \frac{\epsilon^{|-X_{q_i} + X_{q_j} - X_{q_k} + X_{q_l}|}}{\Lambda_F^2} \ .
\end{equation}
where $\Lambda_F$ is the UV scale of the model, and $c^{A}_{ijkl}$ is an unknown dimensionless complex coefficient.  

For FN models that satisfy our global naturalness criterion, it is plausible to expect the different coefficients $c^{A}_{ijkl}$ to be relatively uncorrelated and have $\mathcal{O}(1)$ sizes.\footnote{Support for this assumption can be derived from the near-uncorrelated near-log-normal distributions of the individual SM observables in our numerical scans, see Fig.~\ref{fig:singleParamHistograms} in Appendix~\ref{app:prior}. It is plausible to expect other observables to behave similarly.} 
We can therefore extend our Bayesian-inspired numerical approach to study the phenomenology of the most SM-like FN quark sector models.~\footnote{For a similar approach, applied to find Yukawa-coupling modifications generated by select FN models due to SMEFT operators involving two fermions and additional Higgs fields, see~\cite{Alonso-Gonzalez:2021tpo}.} 

By construction, the most interesting observables in this context are quark flavor-changing observables. 
The implementation proceeds as follows: For each set of flavor charges in Table~\ref{tab:bestWithin2}, we make random choices for the Yukawa coefficients $c^{u,d}$ in Eq.~(\ref{eq:YukawaMatrixInTermsOfCEpsilon}) as 
outlined in Section~\ref{s.numerical}. If a given coefficient choice gives a SM-like quark spectrum  ($\XiSM < 2$), we use it as the starting point to find an exact (and fully natural) SM fit, as detailed in Appendix~\ref{app:smfits}, thereby determining the corresponding $\epsilon$ and rotation matrices $U_{u,d}, W_{u,d}$. 
We then randomly select coefficients for all possible SMEFT operators in the flavor basis, using the same prior distribution as for $c^{u,d}$. 
At this point, the theory is fully specified up to the unknown flavor scale $\Lambda_F$, and for each particular set of Yukawa and dimension-6 coefficient choices, we can perform the rotation to the quark mass basis and the matching to the Low-Energy Effective Theory (LEFT), which is ultimately used to obtain predictions for the observables.  These predictions, as well as the corresponding SM predictions and experimental bounds, are taken from the package \texttt{flavio}~\cite{Straub:2018kue}, in the context of a new version of \texttt{smelli}~\cite{Aebischer:2018iyb,SmolkovicStangl}. Repeating this many times for each texture in Table~\ref{tab:bestWithin2} gives a picture of the ``theoretically preferred" distribution of observables for each texture.

Unsurprisingly, the most stringent constraint on $\Lambda_F$ -- typically around $\sim 10^5~\mathrm{TeV}$ -- is set by the absorptive part of $K-\bar K$ mixing, followed by the absorptive part of $D-\bar D$ mixing (encoded in the observables $\epsilon_K$ and $x_{12} \sin \phi_{12}$, respectively).
This is illustrated in the top row of Figure \ref{fig:phenocorrelations}, where we show the distribution of the flavor scale saturating the bounds on these observables for four representative textures from Table~\ref{tab:bestWithin2}. 
For all such textures, the overall flavor scale is relatively close to the flavor-anarchic expectation. This is because large right-handed rotations between the flavor and mass bases significantly undermine the suppression of flavor-changing interactions present in the flavor basis. Nevertheless, different textures clearly exhibit distinct distributions, especially in the $D - \bar{D}$ direction. 

While it may be possible to observe a signal in $D - \bar{D}$ mixing in the future if the flavor scale is as low as the $\epsilon_K$ bounds permit, distinguishing between different textures would be more plausible if deviations in other flavor observables, such as $B_{d} -\bar{B}_d$ and $B_s -\bar{B}_s$ mixing, were also observed. 
To further illustrate this, we set the flavor scale to saturate the bound on $\epsilon_K$, and obtain predictions for $x_{12} \sin \phi_{12}$ and for the new physics contribution to the $B_{d,s}-\bar{B}_{d,s}$ mixing amplitudes, $\Delta_{\rm NP} m_{B_{d,s}}$. We plot the distributions for $x_{12} \sin \phi_{12}$ and $\Delta_{\rm NP} m_{B_{d}}$ in the bottom row of Figure \ref{fig:phenocorrelations}. (The distribution and experimental reach for $\Delta_{\rm NP} m_{B_{s}}$ is very similar.) The gray band is excluded by current bounds on $D - \bar{D}$ mixing \cite{HFLAVCharmWeb}. The analogous exclusion from $B_d-\bar{B}_d$ mixing is not visible in the plot, the main limit being the current precision in the SM prediction. 
However, we show the current measurement precision on $\Delta m_{B_d}$ as a dashed light blue line \cite{ParticleDataGroup:2024cfk}, to indicate what sensitivity is experimentally plausible if significant theoretical advances were made in computing the SM prediction. 

Similar differences across textures can be observed in other flavor-violating processes, for example rare $B$ and $K$ decays. However, in these cases, the absolute size of the deviations is several orders of magnitude below current bounds, making them experimentally inaccessible in the foreseeable future. For instance, the new physics contribution to $B_s \to \mu^+ \mu^-$ is $7-8$ orders of magnitude below the SM one, well beyond experimental reach. Thus, $\Delta F =2$ observables remain the only viable experimental probes of FN models in the quark sector. These conclusions also hold if the Higgs has a non-zero FN charge, as SMEFT operators involving the Higgs affect these observables only at the loop level.

Our analysis demonstrates how the phenomenology of different FN textures can be understood and potentially used to distinguish between UV models.
In practice, the high flavor scale due to near-anarchic mixing in the right-handed quark sector, and challenges in improving purely hadronic observables both theoretically and experimentally, may make an actual detection of deviations difficult.
However, we anticipate that extending our analysis to FN models of the lepton sector would yield more optimistic experimental predictions for observables involving lepton-flavor violation and possibly simultaneous lepton- and quark-flavor violation. We leave this for future work.

\section{Conclusions}
\label{s.conslusions}

The flavor puzzle remains one of the most profound mysteries in the Standard Model, with the peculiar structure of fermion mass matrices hinting at an underlying principle or symmetry yet to be uncovered. 
Directly probing the mechanisms that generate this structure is difficult, as in most scenarios, tight constraints on flavor violation push the relevant scale of flavor physics to very high values far beyond our ability to probe directly in the intermediate future.
Nevertheless, future experiments may offer the prospect of directly or indirectly probing this elusive flavor structure, making its detection and diagnosis a compelling goal. 

Our work provides a systematic framework for exploring the vast landscape of Froggatt-Nielsen solutions to the quark flavor problem. By significantly expanding the range of viable charge assignments beyond those commonly studied, we present a much more comprehensive picture of FN models. Even so, our global criterion -- requiring a natural generation of the SM mass and mixing hierarchies across the entire parameter space of a model -- remains stringent enough to yield a manageable set of FN models, as presented in Table~\ref{tab:bestWithin2}.

We demonstrated that this  most natural subset of FN benchmark models can generate a range of flavor-violating effects, with varying magnitudes and correlations influenced by the specific flavor charge assignments. In particular, we found that $\Delta F = 2$ observables like $K - \bar{K}$, $D - \bar{D}$, and $B_{d,s} - \bar{B}_{d,s}$ mixing could potentially be used to differentiate between these models if deviations from the Standard Model are observed. This explicitly shows how, in principle, identifying the correct solution to the quark flavor problem could become feasible once flavor-violating signals beyond the SM are unambiguously detected.

Global SMEFT fits that incorporate a specific FN model might be more constraining~\cite{Aoude:2020dwv,Bruggisser:2021duo, Bruggisser:2022rhb,Greljo:2022cah}, and it would be interesting to perform such fits for each of the natural FN models identified in Table~\ref{tab:bestWithin2}, using well-defined priors for the unknown $\mathcal{O}(1)$ coefficients. This approach could further refine our understanding of how different FN textures manifest in low-energy observables and help narrow down viable model candidates.

Our approach to identifying natural FN models can be extended to other sectors, such as FN models for the lepton sector, or more complex frameworks involving multiple or non-abelian flavor symmetries~\cite{Nir:1993mx, Linster:2018avp, Barbieri:1996ww, Barbieri:1995uv, Pomarol:1995xc}. These extensions could uncover additional flavor-violating observables that offer the best prospects for probing and understanding the underlying physics of the flavor puzzle. It may also be interesting to study the impact of our work on dark sectors that are related to the SM via a discrete symmetry~\cite{Chacko:2005pe}, and to revisit our phenomenological analysis if the broken $U(1)_X$ is an approximate global symmetry, leading to a flavon degree of freedom in the spectrum (see e.g.~\cite{Ema:2016ops, Calibbi:2016hwq, Bauer:2021mvw}). We leave such investigations for future work.

\begin{acknowledgments}

It is a pleasure to thank Savas Dimopoulos, Marco Fedele, Christophe Grojean, Gudrun Hiller, Anson Hook, Seyda Ipek, Junwu Huang, Micah Mellors, Patrick Owen, Yael Shadmi, Peter Stangl, Ben Stefanek, and Alan Schwartz for valuable discussions.
CC, DC and JOT would like to thank Perimeter Institute for hospitality during the completion of this work.  This research was supported in part by Perimeter Institute for Theoretical Physics. Research at Perimeter Institute is supported by the Government of Canada through the Department of Innovation, Science and Economic Development and by the Province of Ontario through the Ministry of Research, Innovation and Science.
The research of CC was supported by the Cluster of Excellence \textit{Precision Physics, Fundamental Interactions, and Structure of Matter} (PRISMA$^+$, EXC 2118/1) within the German Excellence Strategy (Project-ID 39083149).
The research of DC was supported in part by a Discovery Grant from the Natural Sciences and Engineering Research Council of Canada, the Canada Research Chair program, the Alfred P. Sloan Foundation, the Ontario Early Researcher Award, and the University of Toronto McLean Award.
The research of ETN was supported by the U.~S.~Department of Energy (DOE), Office of Science, Office of High Energy Physics, under Award Number DE-SC0010005.
\end{acknowledgments}

\appendix

\section{Standard Model Fits and Tuning}
\label{app:smfits}

\begin{table}[]
    \centering
    \begin{tabular}{c|c}
        SM parameter & Value \\
        \hline
$m_u$ & 0.00117(35) \\
$m_c$ & 0.543(72) \\
$m_t$ & 148.1(1.3) \\
$m_d$ & 0.0024(42) \\
$m_s$ & 0.049(15) \\
$m_b$ & 2.41(14) \\
$|V_{12}|$ & 0.22450(67) \\
$|V_{13}|$ & 0.00382(11)\\
$|V_{23}|$ & 0.04100(85) \\
$|J|$ & 3.08(14) $\times 10^{-5}$
     \end{tabular}
    \caption{Standard Model parameters and uncertainties used to compute goodness-of-fit parameters such as $\XiSM$ (Eq.~\ref{eq:defXi}) Quark masses (given in GeV) are taken from \cite{Xing:2011aa}, evaluated at $\mu = 1$ TeV. Running these parameters to the $\mathcal{O}$(PeV) flavor scale would not significantly change our results. CKM parameters are taken from the PDG review \cite{ParticleDataGroup:2024cfk}, evaluated at low scales; we explicitly checked that including the running of the CKM parameters does not have any noticeable effect on our analysis.  
    \label{tab:SMparams}}
\end{table}

In this appendix we present details on finding solutions for the coefficients $c_{ij}^{u,d}$ in Eq.~\ref{eq:YukawaMatrixInTermsOfCEpsilon} that represent realistic SM fits within experimental uncertainties for the quark masses and mixings, see Table~\ref{tab:SMparams}, for each of the FN charge assignments in Table~\ref{tab:bestWithin2}. 

In order to obtain SM fits, we used a simplex minimization algorithm to adjust the coefficients $c_{ij}^{u,d}$ and the parameter $\epsilon$ in order to optimize the standard $\chi^2$ score,
\begin{equation}
\chi^2_{\rm SM} = \sum_i \left( \frac{\mu_i - \mu_i^{\rm SM}}{\sigma_i^{\rm SM}}\right)^2
\end{equation}
where as in Sec.~\ref{s.numerical} the index $i$ runs over all of the SM quark masses, three CKM mixing angles, and $|J|$.  For all of the textures in Table~\ref{tab:bestWithin2}, we are able to find fits with average deviation of less that 2$\sigma$ over all 10 parameters that we fit to; we were further able to find fits with average deviation less than 1$\sigma$ for more than half of the textures in the table, including all of the top 5.  We have further verified that for these fits, the values of the coefficients $c_{ij}^{u,d}$ do not have a significantly different distribution than the prior used to generate the same coefficients for our numerical scans, indicating that the fits do not require any substantial drift away from our $\mathcal{O}(1)$ assumption.
Our final check is to verify that these fits are not locally tuned. 

A standard tuning test is the Barbieri-Giudice measure~\cite{Ellis:1986yg, Barbieri:1987fn}, which for a single SM observable $\mathcal{O}_K$ can be defined as:
\begin{eqnarray}
    \Delta_\mathrm{BG}^K \equiv \max_{k} |\delta_{K,k}| &,&
    \delta_{K,k} \equiv
    \frac{\delta \log \mathcal{O}_K}{\delta \log c_k }\ ,
\end{eqnarray}
where $c_k$ runs separately over the real and imaginary parts of each of the $c_{ij}^{u,d}$ coefficients defined in Eq.~\ref{eq:YukawaMatrixInTermsOfCEpsilon}.  A second maximization then gives the overall tuning taking into account all fitted SM observables:
\begin{eqnarray}
    \Delta_\mathrm{BG} \equiv \max_{K} \Delta_\mathrm{BG}^K \ .
\end{eqnarray}
This basically represents the maximum sensitivity of all the 10 SM observables $\mathcal{O}_K$ with respect perturbing any single one of the 18 complex coefficients $c_{ij}^{u,d}$.
However, we argue that for Froggatt-Nielsen models and their many coefficients, which  arise from some UV theory, this tuning measure is of limited utility, since a small change in the UV theory would slightly change all the coefficients at once, not just one at a time. Indeed, if random small permutations of some characteristic size are applied to \textit{all} $c_{ij}^{u,d}$ coefficients of a given SM fit, we numerically find that $\Delta_\mathrm{BG}^K$ generally significantly \textit{underestimates} the variance of a given SM observable $\mathcal{O}_K$. In other words, for FN models, the Barbieri-Giudice measure underestimates tuning.

Some way of assessing this total sensitivity is required. We therefore define the following tuning measure for each SM observable $\mathcal{O}_K$:
\begin{eqnarray}
    \Delta_\mathrm{tot}^K \equiv \left[ \sum_s (\lambda^K_s)^2 \right]^{1/2}
\end{eqnarray}
where the $\lambda_s^K$ are the eigenvalues of the matrix $\Delta_{kl}^K$ defined as
\begin{equation}
    \Delta_{kl}^K \equiv \frac{\delta^2 \log \mathcal{O}_K}{\delta \log c_k \delta \log c_l}
\end{equation}
and $c_k$ runs over the real and imaginary parts of all $c_{ij}^{u,d}$ coefficients as above.
It is easy to see that this quantity could give a more complete account of the tuning of a given SM fit, since the sum in quadrature over the principal directions of $\Delta_{kl}^K$ takes into account the total variability of SM observable $\mathcal{O}_K$, regardless of whether the direction of maximum sensitivity  is aligned with any one $c_{ij}^{u,d}$, which is appropriate when perturbing all coefficients at once. Indeed, when varying all coefficients by  random perturbations of characteristic relative scale $\sigma$, we find that $\Delta^K_\mathrm{tot} \sigma$ gives a very good direct estimate of the resulting relative variance of $\mathcal{O}_K$. This supports the argument that $\Delta_\mathrm{tot}^K$ is a more faithful measure of sensitivity to changes in the underlying UV theory of a FN model, and an overall tuning measure for the whole SM fit can be obtained by summing in quadrature over all SM observables:
\begin{equation}
    \Delta_\mathrm{tot} = \left[ \sum_K (\Delta_\mathrm{tot}^K)^2\right]^{1/2}
\end{equation}

\begin{table}[]
    \centering
    \begin{tabular}{c|c|c|c|c}
        Rank by this prior & Rank in Tab.~\ref{tab:bestWithin2} & $\mathcal{F}_2 \; (\%)$ & $\mathcal{F}_5 \; (\%)$ & $\epsilon$ \\
        \hline
1 & 1 & 0.17 & 24 & 0.16 \\
2 & 2 & 0.13 & 22 & 0.17 \\
3 & 3 & 0.12 & 19 & 0.12 \\
4 & 16 & 0.11 & 19 & 0.2 \\
5 & 8 & 0.1 & 17 & 0.1 \\
6 & 15 & 0.1 & 18 & 0.21 \\
7 & 18 & 0.1 & 14 & 0.18 \\
8 & 11 & 0.1 & 16 & 0.06 \\
9 & 12 & 0.09 & 19 & 0.21 \\
10 & 10 & 0.09 & 18 & 0.15
     \end{tabular}
    \caption{Analogue of Table~\ref{tab:bestWithin2} for top-10 textures and their $\mathcal{F}_2$, $\mathcal{F}_5$, and average $\epsilon$, but using the ``wider log-normal'' prior for the coefficients of the $c^{u,d}$ matrices.
    For comparison, we also show their rank in Table~\ref{tab:bestWithin2} in the second column.
    The overall fraction of $c^{u,d}$ matrices with this prior that give good fits to the SM is smaller for all textures, which is to be expected because this prior allows for larger random hierarchies in the $c^{u,d}$ coefficients and thus washes out the effect of the texture itself in generating the SM hierarchies.  Textures that are perform well in Tab.~\ref{tab:bestWithin2} also perform well with this prior, although the exact ordering is somewhat shuffled.}
    \label{tab:bestWithin2widerGauss}
\end{table}

\begin{table}[]
    \centering
\begin{tabular}{c|c|c|c|c}
        Rank by this prior & Rank in Tab.~\ref{tab:bestWithin2} & $\mathcal{F}_2 \; (\%)$ & $\mathcal{F}_5 \; (\%)$ & $\epsilon$ \\
        \hline
1 & 1 & 4.0 & 71 & 0.15 \\
2 & 2 & 3.2 & 79 & 0.16 \\
3 & [23] & 0.6 & 40 & 0.14 \\
4 & 13 & 0.4 & 48 & 0.22 \\
5 & 5 & 0.4 & 46 & 0.21 \\
6 & 4 & 0.4 & 58 & 0.14 \\
7 & [22] & 0.4 & 78 & 0.17 \\
8 & [21] & 0.3 & 46 & 0.13 \\
9 & 9 & 0.2 & 62 & 0.14 \\
10 & [28] & 0.2 & 39 & 0.24
     \end{tabular}
    \caption{
    Analogue of Table~\ref{tab:bestWithin2} for top-10 textures and their $\mathcal{F}_2$, $\mathcal{F}_5$, and average $\epsilon$,  but using the ``uniform flat'' prior distribution for the coefficients of the $c^{u,d}$ matrices.
    For comparison, we also show their rank in Table~\ref{tab:bestWithin2} in the second column. The textures with rank in [square brackets] do not appear in Table~\ref{tab:bestWithin2}, but we show their rank if that table was continued to the top-30. These textures have the following charges: rank 3 (this table), \{3, 2, -3, -2, -3, -3, -3\}; rank 7, \{3, 2, -4, -2, -4, -4, -3\}; rank 8, \{3, 2, -3, -2, -3, -3, -2\}; rank 10, \{4, 3, -4, -2, -4, -4, -4\}.]
    } 
    \label{tab:bestWithin2flat}
\end{table}

We find that it is generally easy to find SM fits for the top-20 textures in Table~\ref{tab:bestWithin2} that have $\Delta_\mathrm{tot} \lesssim $ a few, proving that these solutions to the SM flavor problem are truly untuned with respect to the underlying details of the UV theory. Informally, we notice that there seems to be a trend that the $\Delta_\mathrm{tot}$ of an average SM fit increases modestly for lower-ranked textures in Table~\ref{tab:bestWithin2}. While further study would be required to solidify this relationship, it provides further suggestive evidence that our global naturalness criterion of ranking textures by their $\mathcal{F}_2$ or $\mathcal{F}_5$ fractions is the fundamental measure of how SM-like a texture wants to be, and how natural any SM fits are that do exist.

\begin{figure}[t]
\centering
\includegraphics[width=\linewidth]{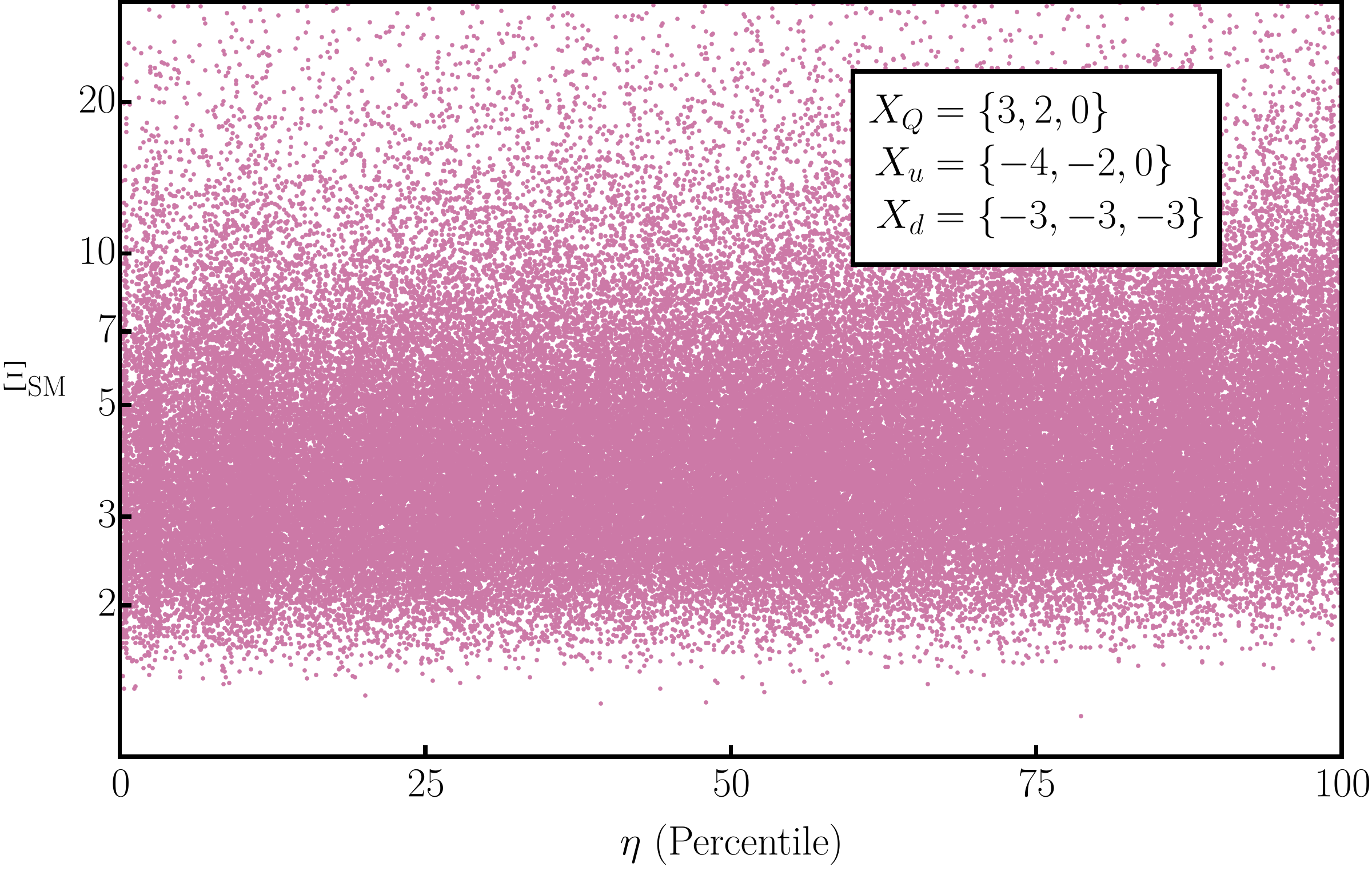}
\caption{Correlation between closeness to the SM ($\XiSM$) and how anomalous the random hierarchy between the largest and smallest $c_{ij}^{u,d}$ coefficients is (anomalous here means a particularly low or high $\eta$ pecentile), for the example of the top texture in Table~\ref{tab:SMparams}.  The lack of clear correlation demonstrates that the SM-like hierarchies do not rely on accidental hierarchies amongst the coefficients in Eq.~\ref{eq:YukawaMatrixInTermsOfCEpsilon}.}
\label{fig:XietaCrossCorrelation}
\end{figure}

\begin{figure}[t]
 \centering
 \includegraphics[width=\linewidth]{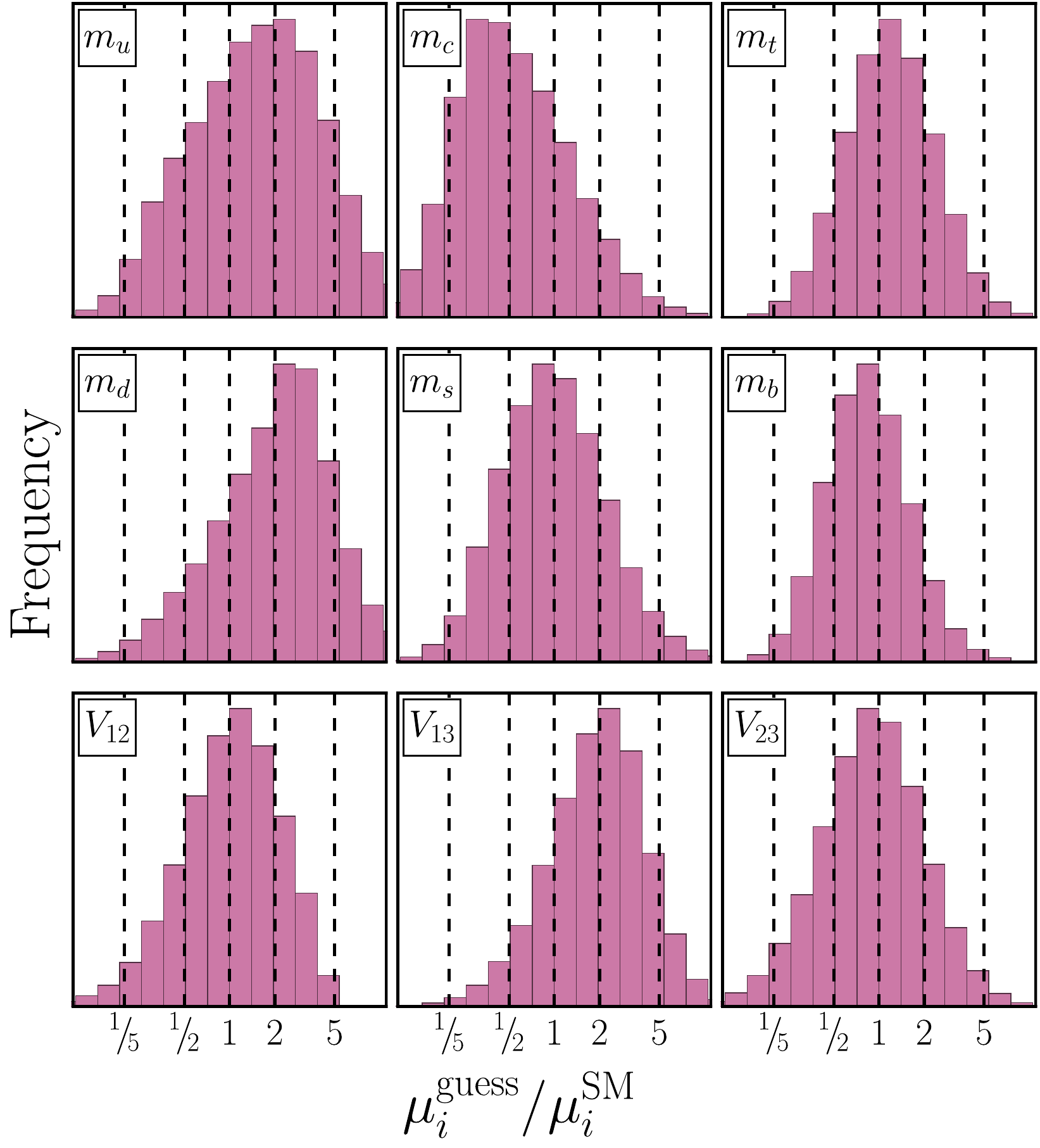}
 \caption{
    Individual parameter distributions for the best texture listed in Table~\ref{tab:bestWithin2}.  For the quark masses and the three independent CKM elements, each histogram shows the distribution of one particular parameter relative to its true SM value.  For this texture, all parameter distributions are peaked within a factor of a few from their SM values.  Although not shown here, there is minimal cross-correlation between any pair of these parameters.}
    \label{fig:singleParamHistograms}
\end{figure}

\section{Dependence on prior distributions and statistical checks} \label{app:prior}

In this appendix we collect the details of the prior distributions and statistical measures we use in our analysis.  Because one of the main products of this paper is a collection of ``statistically good'' textures, i.e.\ a collection of textures which do a good job of reproducing SM-like hierarchies for somewhat arbitrary $\mathcal{O}(1)$ Yukawa coefficients $c^{u,d}$, it is important to understand how much these results depend on our priors for these coefficients.  While we do find that the precise numbers claimed here depend on this distribution, our derived  list of good textures is robust with respect to the exact choice of prior for the coefficients, up to some minor reordering. 

For all results shown in the main body of this paper, our numerical scan generated the entries of $c^{u,d}$ as follows: for each coefficient independently, a magnitude was drawn from a log normal distribution centered around 1 with a standard deviation of $\ln 10^{0.3}$ (to enforce that all coefficients in the matrix be $\mathcal{O}(1)$), and a phase was drawn from a flat distribution between $0$ and $2 \pi$.  This choice yields the results shown in Table~\ref{tab:bestWithin2}, namely that there are a few textures for which $\mathcal{O}(2\%)$ of randomly generated coefficients yield quark masses and CKM elements within a factor of 2 of their SM values, and $\mathcal{O}(50\%)$ yield parameters within a factor of 5.

In order to check that this list is robust to other choices of distribution, we ran two other scans for all charge assignments:
\begin{enumerate}
\item A ``wider log-normal'' scan where we drew the $c^{u,d}$ coefficients with a uniform distribution in phase and a log-normal distribution in magnitude centered on 0 and with a standard deviation of $\ln 10^{0.6}$.
\item A ``uniform flat'' scan where we drew the real and imaginary parts of each entry of $c^{u,d}$ separately from uniform flat distributions between $-3$ and $3$.
\end{enumerate}

The lists of the top 10 textures (ranked by the fraction $\mathcal{F}_2$ within a factor of 2 of the SM value) for each of these scans are given in Tables~\ref{tab:bestWithin2widerGauss} and~\ref{tab:bestWithin2flat}.  We can see that the rough list and ordering of textures in the top $5$ is not significantly changed by any of these choices, so we conclude that this data is  robust to changes of distribution (provided all entries of $c^{u,d}$ be $\mathcal{O}(1)$).

Another reasonable concern that one may have about these results is whether the observed hierarchies are primarily coming from the texture choice.  Our prior distribution for $c^{u,d}$ is chosen to result in $\mathcal{O}(1)$ coefficients, but there can still be anomalously hierarchical draws from this distribution, and it is possible that there are textures in which it is precisely these anomalously hierarchical draws that give results closer to the SM.  To test this, we construct a measure of how hierarchical a particular choice of coefficients $c^{u,d}$ is with respect to a given prior distribution.  Namely, we define
\begin{equation}
    \eta \equiv \log_{10} \frac{\max_{i,j} | c^{u/d}_{i j} |}{\min_{i,j} | c^{u/d}_{i j} |}.
\end{equation}
Given a distribution on the individual coefficients we can construct the distribution of $\eta$ and find a measure of how anomalously hierarchical a particular choice of $c^{u,d}$ are.  For the textures displayed in Table~\ref{tab:bestWithin2}, there is generally very little correlation with how anomalously high or low $\eta$ is and how good of a fit the coefficients are to the SM (quantified by the smallness of $\XiSM$).  For example, the cross-correlation of these two parameters for our best texture is shown in Fig.~\ref{fig:XietaCrossCorrelation}.  We take this as evidence that the observed hierarchies in the good textures are driven almost entirely by the textures themselves rather than the Yukawa coefficient matrices.

For a given texture, we can also construct histograms of the individual quark masses and CKM elements to verify that the texture predicts all of them to be approximately their SM values.  We do this by generating many random choices of $c^{u,d}$, varying $\epsilon$ to minimize $\XiSM$ for each choice of $c^{u,d}$, and then plotting the resulting distributions of the quark masses and CKM elements.  As an example, in Fig.~\ref{fig:singleParamHistograms} we show the distributions for the individual parameters for our best-performing texture in Table~\ref{tab:bestWithin2}.  We have checked these distributions for all textures listed in Table~\ref{tab:bestWithin2} and we find that all parameter distributions are centered within a factor of a few of the true SM values.  We have also checked the cross-correlations between these parameters for the best textures listed in Table~\ref{tab:bestWithin2}, and they are relatively mild, meaning that each parameter behaves approximately as if it were drawn independently from a distribution of the type shown in Fig.~\ref{fig:singleParamHistograms}.

\bibliography{bibliography}

\end{document}